\documentclass[
twocolumn,
amsmath,amssymb,
aps, physrev,
pra,
]{revtex4-2}
\usepackage{lipsum, babel}
\usepackage[utf8]{inputenc}
\usepackage{graphicx}
\usepackage{mathrsfs}
\usepackage{placeins}
\usepackage{rsfso}
\usepackage{hyperref}
\usepackage{amsmath}
\usepackage{subcaption}
\usepackage{dcolumn}
\usepackage{bm}
\usepackage{ragged2e} 
\usepackage{xcolor}
\usepackage{float}   
\usepackage{ulem} 
\begin{document}

\title{\textbf{Collective Chiral Response of Achiral Plasmonic Arrays Induced by Truncated Bloch-Floquet Effect}
}
\author{
Sahil Sahoo$^{1}$,
Ahmed Lafeef Ettapuram Naduvilepurayil$^{1}$,
Akash Das$^{1,2}$,
Yuri Gorodetski$^{1,2,*}$
}
\affiliation{$^{1}$Department of Electrical and Electronic Engineering, Ariel University, Ariel 40700, Israel,}
\affiliation{$^{2}$Department of Mechanical Engineering and Mechatronics, Ariel University, Ariel 40700, Israel.}
\email{Contact author: yurig@ariel.ac.il}

\begin{abstract}

Chirality in plasmonic systems is commonly associated with structures differentiated by their handedness. Here, we demonstrate the emergence of collective chirality in \textit{structurally achiral} periodic hole arrays induced by non-local surface-plasmon interactions. We attribute this effect to the transfer of the optical angular momentum in the system. While an individual sub-wavelength circular aperture is not expected to alter the total angular momentum,
the periodic structure results in a coherent coupling of plasmons wavefronts, giving rise to a collective Bloch-Floquet-like response. The edge of the structure accumulates the helical phase that eventually results in a visible spin-dependent shifting of the plasmonic beam. We systematically study the effects of hole-array size, periodicity and the geometry. The observed phenomenon shows an unprecedented robustness and consistency. These results establish periodic plasmonic arrays as a platform for converting localized helicity-dependent responses into collective chirality, providing a route toward controlling chiral plasmonic wavefronts in structurally achiral systems.

\end{abstract}
\maketitle

\section{Introduction}
The interaction of circularly polarized (CP) light with plasmonic nanostructures has emerged as a powerful platform for manipulating optical chirality at sub-wavelength scales \cite{berova2000circular,maier2007plasmonics,barron2009molecular,sun2014spinning,meinzer2014plasmonic,lininger2023chirality,ettapuram2025direct,sahoo2026spatio,li2026hyperbolic}. Chirality is a property of a structure or a field existing in two enantiomeric states inter-convertible by a mirror symmetry. Optical response in chiral systems depends strongly on the handedness of the incident circularly polarized light, producing distinct pattern signatures for opposite helicities \cite{tullius2015superchiral,sun2020selectively}. Chiral plasmonic nanostructures have attracted considerable attention due to their ability to generate strong circular dichroism \cite{gibbs2013plasmonic,wang2017circular,zu2019imaging,hu2022plasmonic} and enhanced chiral electromagnetic fields. This motivated numerous applications in enantioselective sensing, chirality detection, and engineered light–matter interactions \cite{valev2013chirality,luo2017plasmonic,hentschel2017chiral,wang2025chiral, anker2008biosensing, hendry2010ultrasensitive, schaferling2012tailoring}. 

A broken rotational symmetry in a structure leads to the exchange of angular momentum with the incident optical field, so that the total optical angular momentum is not conserved. A circular hole in a conducting thin film has been recently presented as a system maintaining the total optical angular momentum thanks to its continuous rotational symmetry. It has been shown that the spin angular momentum (SAM) of circularly polarized light can be coupled to the orbital angular momentum (OAM) through spin-orbit interaction, resulting in a helicity-dependent surface plasmons (SPs) field \cite{bliokh2015spin,aiello2015transverse,gorodetski2013generating,spektor2021orbital,fu2024transverse,sahoo2026revealing}. Nevertheless, an isolated hole retained an achiral in momentum space ($k$-space) response, a phenomenon associated with hidden or local chirality \cite{hashiyada2014local,davis2013superchiral,zu2018deep,horrer2020local,cen2024chirally}. 

Often, periodic arrays of plasmonic holes are used to resonantly excite  SPs by a specific incident field \cite{zayats2005nano,bouillard2014shaping}.  While each hole can be regarded as a spin-dependent plasmonic source with a well-defined phase, the periodic arrangement establishes a fixed spatial relationship between neighboring sources. This periodic coupling and interference can be described in terms of collective Bloch–Floquet modes \cite{faisal1997floquet,bilitewski2015scattering}, in which the field acquires a well-defined phase upon translation by a lattice vector. Thus, the periodicity does not merely replicate the local chiral response, but provides the framework for its coherent propagation and collective buildup across the array. Consequently, the field structure is no longer confined to the vicinity of individual holes but can emerge as a global azimuthal phase along the array, giving rise to a collective nonlocal chiral response. Illuminating a square hole array by a collimated optical beam results in a plasmonic quasi-planewave launched at the structure edge. Here, the global phase acquired in the collective response of the coupled holes may leave a considerable signature in the real-space ($r$-space) distribution and result in unexpected beam shifts. Therefore, the illumination of the achiral lattice with opposite circular-polarization handedness produces two distinct chiral $r$-space configurations. The corresponding $r$-space shifts reflect phase-dependent interactions that are manifested in the $k$-space, making the chirality observable in both $r$- and $k$-space. This provides direct evidence of nonlocal chirality emerging from an achiral structure. Notably, increasing the size of the array (by increasing the number of holes) strengthens the non-local character of the phase and diminishes the local contribution of a single source. Accordingly, the aforementioned beam-shifts or tilts become less significant with respect to the beam width. 

In this Letter, we introduce the concept of non-local chirality in periodic arrays of structurally achiral plasmonic holes and demonstrate how collective interactions can transform a localized helicity-dependent response into a measurable chiral state extending across the lattice. By systematically varying the lattice size, we demonstrate that increasing the number of plasmonic sources progressively strengthens their coherent interaction, revealing a transition from predominantly local behavior to a collective non-local response. As the array becomes sufficiently extended, these collective modes enhance the non-local contribution and give rise to pronounced chiral features in both the real and momentum space. 

We investigate the effect of the lattice periodicity, rotation symmetry and SP propagation length on the observed phenomenon. We confirm that a circular-polarization helicity and a local spin-dependent excitation alone are insufficient to produce an observable macroscopic chiral response. The emergence of collective chirality is governed by the interplay between lattice periodicity, array size, plasmonic coupling, and the phase coherence of the excited modes. Our results establish a direct connection between local spin–orbit-induced chirality, Bloch–Floquet-mediated collective interactions, and non-local chiral phenomena in structurally achiral plasmonic metasurfaces. This provides a general framework for understanding and controlling when chirality can emerge, propagate, and become observable in periodic plasmonic systems, opening new opportunities for chiral wavefront engineering and the development of next-generation plasmonic metamaterials, metasurfaces, and chiral nanophotonic devices.

\section{Collective Chirality Observation}
\begin{figure*}
    \includegraphics[width= 1\textwidth]{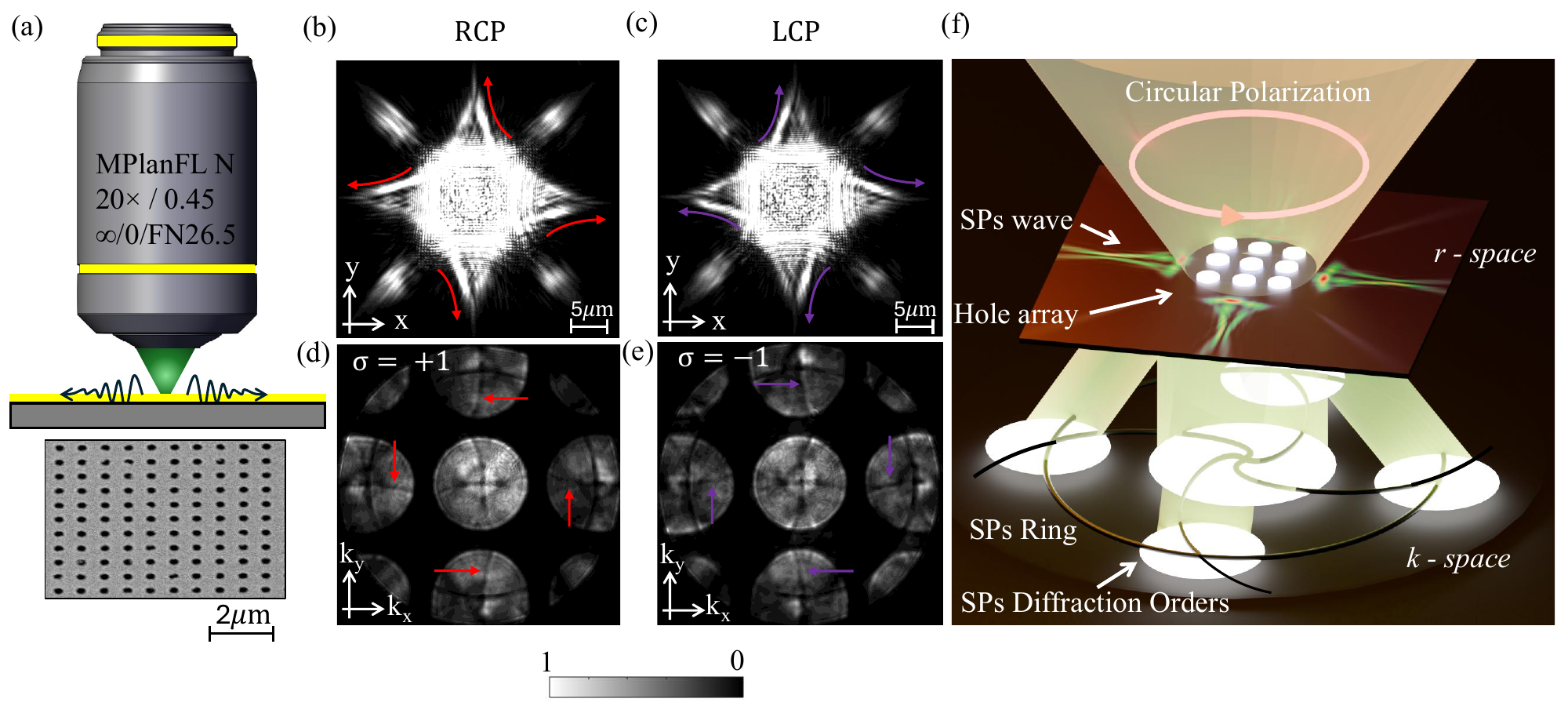}
    \caption{\label{schematic chirality} \justifying
    (a) Schematic illustration of the focusing geometry for the metallic hole array and the scanning electron microscope (SEM) image of the square hole-array. (b,c) Experimentally measured $r$-space plasmonic-wave interference patterns under RCP and LCP illumination, respectively. (d,e) Corresponding $k$-space results for RCP and LCP illumination. (f) Schematic illustration of the plasmonic-wave propagation and the resulting swirling intensity distribution associated with the chiral response.}
\end{figure*}

The SPs are excited at the metal–dielectric interface as shown in Fig. \ref{schematic chirality}. Additional details of the experimental setup are shown in supporting information (S1). As the SPs propagate away from the array, the incident light's handedness becomes clearly observable through the spatial distribution of their wavefronts. Under CP illumination, the generated plasmonic fields exhibit a clear directional asymmetry, with one plasmonic focusing arm appearing significantly brighter than the opposite one. Upon switching the incident helicity from right to left circularly polarized (RCP/LCP) this asymmetry reverses, interchanging the relative intensities of the two plasmonic beam arm, as shown in Fig. \ref{schematic chirality}b,c for the $\Lambda = 765$ nm period hole array (the blue and the red arrows mark the dominating arm). As shown in Fig. \ref{schematic chirality}d,e the helicity dependent SPs distribution can be also observed in the $k$-space by using a dedicated Fourier lens (see S1 for the details). A prominent SP ring is observed with radius $k_{sp}$, where $k_{SP} = k_0\sqrt{\frac{\epsilon_m}{\epsilon_m+\epsilon_d}}$, where $k_0 = 2\pi/\lambda_0$, $\epsilon_m$ and $\epsilon_d$ are the free space incident wavenumber, the dielectric constant of the metal (Au) and the dielectric constant of the substrate (SiO), respectively. The SP excitation is achieved by the overlap of the diffraction orders (seen as full circles) with a fraction of the SP ring. The high NA illumination makes these orders rather large generating a wide spatial frequency spectrum in the plasmonic beam. The primary SP resonance is replicated into all the diffraction orders by the Bloch-Floquet theorem. The interference between these SP rings gives rise to characteristic chiral intersection patterns in the $k$-space. This behavior reveals the appearance of a helical phase in the plasmonic wave front and may be linked to the transfer of the incident spin angular momentum to the SP field. Fig. \ref{schematic chirality}f schematically depicts this process. While the utilized square hole-array does not exhibit any built-in structural chirality our analysis focuses on the parameters of the illuminating beam. The continuous imprinting of CP across the free-space wavefront generates a spatially varying phase structure. Upon focusing, the resulting longitudinal field component inherits this phase, which is subsequently transferred to the plasmonic excitation through coupling at the metallic interface. As a result, the excited plasmonic field experiences spatial spin-dependent shifting.

Consider a uniformly circularly polarized plane wave propagating along the $+z$-axis, $ \mathbf{E}_{\mathrm{in}}(r) = E_0(r)\!\left(\hat{\mathbf{x}} + i\sigma\hat{\mathbf{y}}\right)$, where $\sigma$ (SAM) $= \pm 1$ is the spin number $\sigma = +1$ for RCP, $\sigma = -1$ for LCP and $E_0(r)$ is the radial amplitude envelope. The inherent cylindrical symmetry demands expressing the field in a local frame that rotates with the azimuthal angle, $\phi$ at each point in the transverse plane. Using the standard geometric identities: $\hat{\mathbf{x}} = \cos\phi\,\hat{\mathbf{r}} - \sin\phi\,\hat{\boldsymbol{\phi}}$, $\&$ $\hat{\mathbf{y}} = \sin\phi\,\hat{\mathbf{r}} + \cos\phi\,\hat{\boldsymbol{\phi}}$, the incident field becomes $\mathbf{E}_{\mathrm{in}}(r,\phi) = E_0(r)\,e^{i\sigma\phi}\!\left(\hat{\mathbf{r}} + i\sigma\hat{\boldsymbol{\phi}}\right)$.

The factor $(\hat{\mathbf{r}} + i\sigma\hat{\boldsymbol{\phi}})$ confirms that the field remains locally circularly polarized at every point with an explicit helical phase $e^{i\sigma\phi}$. The phase is winding continuously and uniformly across the transverse plane with a handedness set by $\sigma$. This demonstrates the conversion of the incident light’s SAM into OAM in the transverse field, with the resulting topological charge given by the spin number $\sigma$. 
The SPs are strictly Transverse Magnetic (TM) modes whose excitation is governed predominantly by the out-of-plane field component $E_z$. 
This longitudinal field can be assessed from the Gauss’s law, \(\nabla \cdot \mathbf{E}=0\), as
\begin{equation}
E_z(r,\phi)\approx \frac{i}{k_z}\frac{\partial E_0(r)}{\partial r}e^{i\sigma\phi} \equiv iA(r)e^{i\sigma \phi},
\end{equation}
where \(k_z\) is the wave number along propagation direction of incident light.
In the paraxial limit, the beam is nearly a plane wave, its wavevectors are almost parallel to $z$, and the transverse spatial gradients of $E_0(r)$ are gentle which results in a negligibly small $E_z$ field. Although the helical wavefront is present in the free space, the coupling with the plasmonic mode is very weak.

However, the longitudinal field becomes more significant as the transverse field varies more rapidly in space. This typically occurs under tight focusing with a high-numerical-aperture (NA) objective. Importantly, the field retains the same azimuthal phase dependence as the transverse component, thereby preserving its phase-winding structure wherever the longitudinal field is nonzero. The NA controls only the amplitude $A(r)$: higher NA means steeper $\partial E_0/\partial r$ and a stronger longitudinal projection, but the phase structure remains untouched by focusing. Tight focusing doesn't create this helical phase but merely transfers it from the transverse field into the longitudinal channel, allowing TM plasmonic modes to access it. With a continuous  wavefront $E_z \propto e^{i\sigma\phi}$ established across the entire focal plane, the hole array fraction illuminated by it adopts the pre-formed phase structure.

\section{Numerical Model}
\begin{figure}[t]
    \centering
    \includegraphics[width=1\columnwidth]{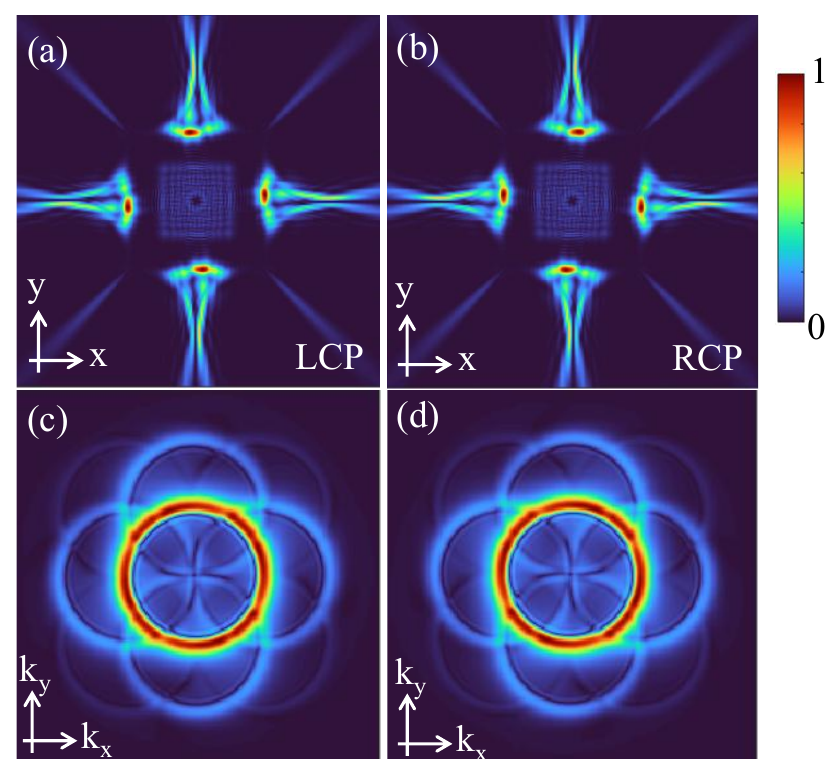}
    \caption{\label{fig:chirality}\justifying (a) Numerical $r$-space intensity distribution calculated for LCP illumination and (b) under RCP illumination. (c) and (d) are the numerical result of the $k$-space distributions for LCP and RCP.}
\end{figure}

To understand the spin-dependent excitation of SPs by a sub-wavelength hole array, we consider a metallic structure consisting of $(N_x \times N_y), N_x = N_y = N$ holes arranged on a square lattice with period $d$. Each hole is treated as a point source of surface plasmons excited by the local incident optical field. The incident beam is assumed to be a focused TM beam with a Gaussian envelope,

\begin{equation}
\begin{aligned}
E_{\mathrm{inc}}(x,y) = {} & \exp\!\left(-\frac{x^2+y^2}{w_0^2}\right) \\
& \times \exp\!\left[ i\!\left( \sigma\phi - \frac{k_{sp}}{2f}(x^2+y^2) \right) \right]
\end{aligned}
\end{equation}
where $\phi = \tan^{-1}\!\left(\frac{y}{x}\right)$, $w_0$ is the beam waist, $k_{sp} = 2\pi/\lambda_{sp}$ is the plasmonic wave number, and $f$ is the focal length. The excitation amplitude associated with the hole located at $(x_m, y_n)$ is obtained from the local value of the incident field, $E_h^{mn} = E_{\mathrm{inc}}(x_m, y_n)$. Therefore, each hole acquires both the amplitude and phase of the incident beam at its position. The spiral phase term $\sigma\phi$ imposes an azimuthally varying phase distribution across the array due to the focusing, while the Gaussian envelope determines the excitation strength of each plasmonic source. The SP field launched from each hole is modeled as a cylindrical surface wave,
\begin{equation}
  G^{mn}(R) = \frac{\exp(i k_{\mathrm{sp}} R)\,\exp(-R/L)}{\sqrt{R}},  
\end{equation}
 
where $R = \sqrt{(x - x_m)^2 + (y - y_n)^2 + f_{obs}}$, and $f_{\text{obs}}$ is the vertical distance between the hole plane and the observation plane, $k_{\mathrm{sp}}$ is the surface-plasmon wave vector and $L$ is the plasmon propagation length. The factor $1/\sqrt{R}$ accounts for the cylindrical spreading of the SP wave. Treating each hole as a secondary plasmonic point source, the total scattered field arises from the coherent superposition of these individual contributions from all the holes, 
\begin{equation}
    E(x,y) = \sum_{m=1}^{N_x} \sum_{n=1}^{N_y} E_h^{mn}\ G^{mn}(R)
\end{equation}
The observable scalar intensity distribution is then calculated from the squared modulus of the total complex field, $I(x,y) = \left|E(x,y)\right|^{2}$, the numerical result for two different helicity have been shown in Fig. \ref{fig:chirality}a,b.

The experimental observation of the $k$-space is achieved by focusing the incident beam to a small point in the center of the hole-array. The plasmonic modes, replicated by the periodicity of the reciprocal lattice, become visible within the area defined by the incident NA, as can be seen in Fig. \ref{schematic chirality}d,e. In order to simulate the $k$-space numerically we define the plasmonic point spread function (PSF) as $C(r) = \frac{e^{ik_{sp}r}}{\sqrt{r}}$, where $r = \sqrt{x^2 +y^2}$. The hole lattice is modeled here by an array of Gaussian distributions decorated by local complex amplitude $E_h$, 
\begin{equation}
\begin{aligned}
    M(x,y) = {} & \sum_{m}\sum_{n} E_{h} \\
    & \times \exp\left(-\frac{(x-x_m)^2 + (y-y_n)^2}{w_{\mathrm{hole}}^{2}}\right),
\end{aligned}
\end{equation}

Taking the Fourier transform of $M(x,y)$ yields $\tilde{M}(k_x,k_y)$, which essentially represents the reciprocal lattice of the metasurface, modulated by the $k$-space distribution of the incident beam.
The total $k$-space field distribution is then obtained by convolving the reciprocal lattice with this fundamental SP ring:
\begin{equation}
\begin{aligned}
    \tilde{E}(k_x,k_y) = \tilde{M}(k_x,k_y) \ast \tilde{C}(k_{sp}).
\end{aligned}
\end{equation}

This convolution represents the physical effect of the Bloch-Floquet theorem. The Fourier transform of the plasmonic PSF, $\tilde{C}(k_{sp})$, generates a single plasmonic ring at radius $k_{sp}$. The convolution then shifts and duplicates the primary SP mode according to the reciprocal lattice $\tilde{M}(k_x,k_y)$, efficiently reproducing the complex Bragg-scattered patterns as shown in Fig. \ref{fig:chirality}c,d for different helicities, We have calculated the SPs ring interference using an alternative method to verify this result S2. Here, the plasmonic ring carries a winding phase characterized by the sign of $\sigma$ induced by the tight focusing of the incident light. The interference between the rings of the different diffraction orders gives rise to the asymmetric intensity pattern observed around the central diffraction order. Noteworthy, the finite width of the plasmonic mode rings is naturally linked to the plasmonic propagation length $\Delta k_r \sim \frac{1}{L}$. We believe that the asymmetry, observed in the plasmonic beam propagation is resulted from the interference of the adjacents plasmonic rings within their overlap area. Accordingly the points of intersections are of special interest and the finite width of the mode lines are crucial for the observation of the effect. In the following section we wish to investigate different scenarios affecting the collective behavior of the structure.

\section{From local to collective chirality}
\begin{figure*}
    \centering
    \includegraphics[width=1\linewidth]{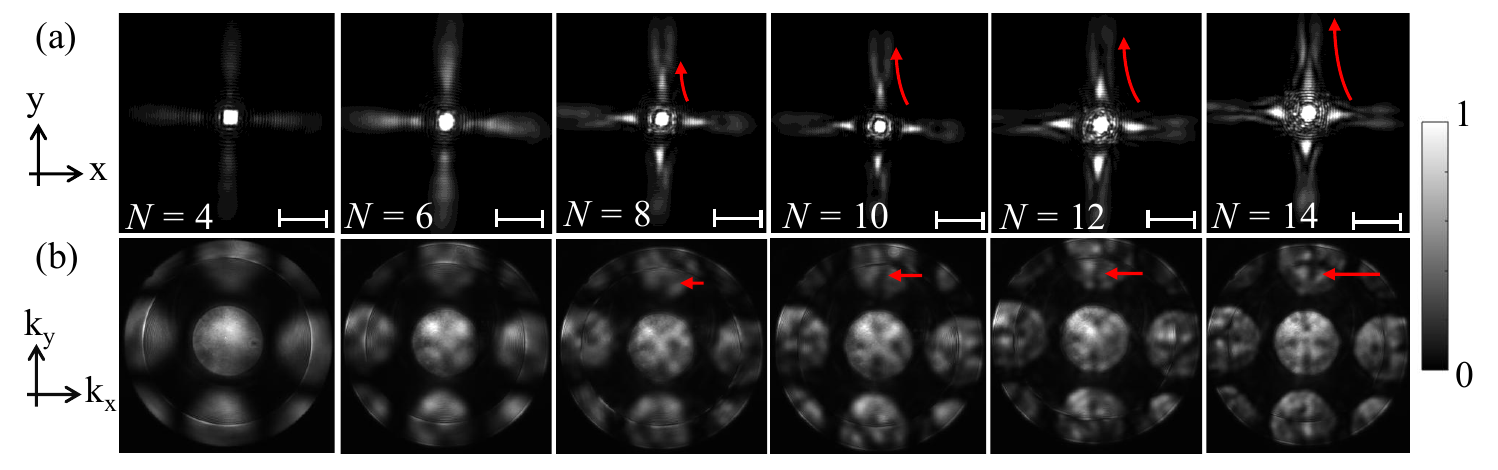}
    \caption{\label{fig:holes variation}\justifying Variation of the optical response with increasing number of holes in the array: (a) $r$-space and (b) $k$-space intensity distributions. Scale bars represent $5 \mu m$ across all panels.}
\end{figure*}
In order to investigate the transition from local to a collective behavior we study the chiral optical response versus the array size. We vary the array size by changing the number of rows/columns in it ($N_X \times N_Y$; $N_X = N_Y = N$). 
\begin{figure}
    \centering
    \includegraphics[width=1\columnwidth]{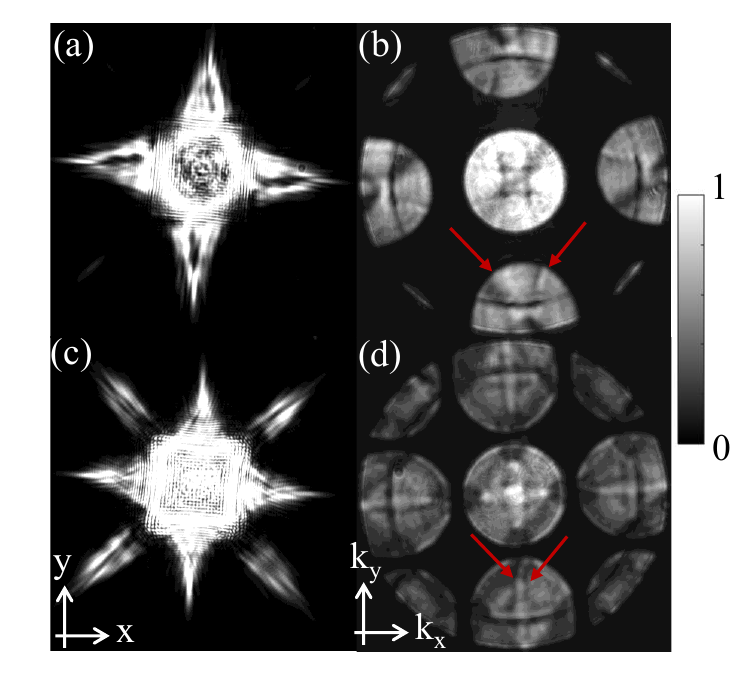}
    \caption{\label{fig:period variation}\justifying $r$ and $k$-space experimental results for hole arrays with different periods under RCP illumination (for LCP refer S6). (a) $r$-space intensity distribution for a period of 705 nm and (b) corresponding $k$-space distribution. (c) $r$-space intensity distribution for a period of 825 nm and (d) corresponding $k$-space distribution.}
\end{figure}
Experimentally, we investigate square arrays with  $N=4,6,8,10,12$, and 14, and examine their intensity distributions in both the $r$-space and the $k$-space. 
In order to ensure similar illumination conditions all the arrays were illuminated with the beam of the same size and same NA. In all cases the spot size was included within the array to ensure beam-like SP distribution. The incident polarization state was chosen to be right-handed ($\sigma = +1$).
As shown in Fig. \ref{fig:holes variation} a and b respectively, the evolution with array size provides a direct visualization of the emergence of chirality. For $N=4$, the intensity distribution remains nearly symmetric, indicating the local response domination accompanied by weak non-local chiral contribution. With increasing $N$, the interaction of SP waves launched by the individual holes becomes increasingly significant, resulting in a progressively stronger asymmetric intensity distribution (marked by the arrow). This can be seen in the $r$-space (Fig.\ref{fig:holes variation}a) as a progressively growing shift of the beam centroid to the left. By switching the polarization to $\sigma = -1$ an equivalent shift to the right appears (not shown here).
In the results shown in the Fig.\ref{fig:holes variation}b, we observe that, as the number of holes in the array increases, the SPs response in the $k$-space indeed evolves from a predominantly local  towards a more collective, Bloch–Floquet-like behavior. For small arrays, only a limited number of holes contribute to the SP field, resulting in fewer interfering plasmonic waves and a weaker collective response. This is seen as a very blurred plasmonic resonant lines. As the array size increases, the proliferation of interacting SP sources intensifies coherent interference, resulting in sharper and more pronounced plasmonic modes. Thus, increasing the array size enables the finite structure to better approximate the collective response of an extended periodic lattice, thereby enhancing the non-local contribution and the resulting chiral response.
In the array with $N=14$, the chiral asymmetry is clearly established, demonstrating the emergence of a collective response. Here we restrict the experimental study to $N=14$ to illustrate the onset and development of the chiral response (the corresponding systematic numerical analysis is provided in the S3).

The observed asymmetry can be referred to the accumulating helical phase at the edge of the array. This phase ramp results in the angular shifting of the SP beam and also manifested in the $k$-space. Intuitively, this phenomenon can be described as a plasmonic nanovortex lattice whose vorticities cancel out within the array but impose a non-zero phase at its boundaries. Accordingly, we suspect that the observed chirality has topological origins. 
To test the effect of the boundary we performed an additional study with very large arrays ($N = 50,60,70$). In this case, the plasmonic wave is expected to significantly decay before reaching the array edge. This represents an ``infinite'' structure that has no edges, therefore the Bloch-Floquet theorem imposes a full phase and amplitude periodicity. The results (shown in the S4) indeed indicate a notable weakening of any asymmetry observed in the SP waves as the structure grows. Additionally, we break the constraints of Bloch-Floquet periodicity by randomizing the hole array. Decoupling the system in this way yields a random interference pattern, completely eliminating any chiral signature in both $r$-space and $k$-space (see S5).

To further investigate the role of collective SP excitations, we systematically change the period of the hole array keeping the other structural parameters unchanged.  
The effect of the period size can be comprehensively visualized by observing the $k$-space. The shorter period expands the diffraction pattern so that the horizontal plasmonic rings do not cross each other. On the other hand the larger period brings the rings closer together and they cross in two points. The resonant period of $\Lambda = 765$ nm brings the horizontal and the vertical modes to cross exactly in the center of each order. The azimuthal helical phase decorating each of the plasmonic rings results in a non-trivial interference in these cross-points. Finally, the pattern appearing in each of the diffraction orders modifies the primary plasmonic resonance and apparently leads to the asymmetry observed in the $r$-space.  

The measured results are shown in Fig. \ref{fig:period variation}a,b for the small period ($\Lambda = 705$ nm) and in Fig. \ref{fig:period variation}c,d for the large period ($\Lambda = 825$ nm). The crossing secondary plasmonic modes are marked by the arrows. The asymmetry effect can be still traced in the $r$-space images however with a decisively weaker magnitude than the resonant structure in Fig.\ref{schematic chirality}.
These observations demonstrate that the chiral response is governed by a combination of the spatial SP modes arrangement in the $k$-space controlled by the period and the NA effect of the incident beam. In the square hole array ($C_4$ symmetry) the maximum effect is achieved with the resonant period. 
\begin{figure}
    \centering
    \includegraphics[width=1\columnwidth]{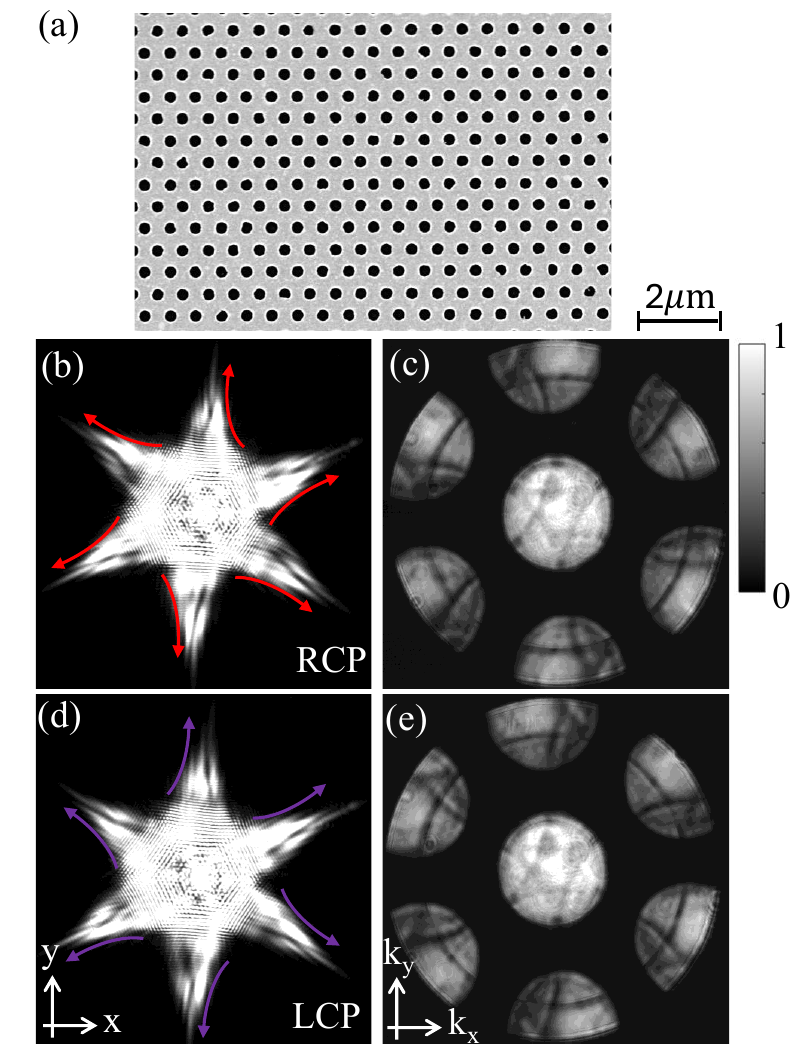}
    \caption{\label{fig:hexagonal array} Experimental results for a hexagonal hole array illuminated with circularly polarized light. (a) SEM image of the hexagonal hole-array, (b) $r$-space and (c) $k$-space intensity distributions under RCP illumination. (d) $r$-space and (e) $k$-space intensity distributions under LCP illumination.}
\end{figure}

Further, we examine the observed chirality dependence on the rotational symmetry of the plasmonic array by changing it from ($C_4$) to ($C_6$). This allows us to address the importance of particular rotational symmetry. As shown in Fig. \ref{fig:hexagonal array}a SEM image of hexagonal ($C_6)$ arrangement of holes is illuminated with circularly polarized light of opposite handedness, ($\sigma=\pm1$) in Fig. \ref{fig:hexagonal array}b,c. The resulting plasmonic beams exhibit opposite handedness for the two illumination states, demonstrating that the chirality is not only preserved but becomes even more pronounced when increasing the rotational symmetry. We noted, however, that in the array with $C_6$ symmetry the strongest chiral signature is obtained with a non-resonant period of $\Lambda = 765$ nm as compared to the resonant array with $\Lambda = 856$ nm (shown in the S7). 

In the $k$-space, the interference of the plasmonic modes produces a characteristic asymmetric intensity distribution around the first-order diffraction spots as shown in Fig.\ref{fig:hexagonal array}b and d, whose handedness reverses upon the reversal of the incident polarization. Although the ($C_4\rightarrow C_6$) transformation modifies the diffraction geometry, the spin-dependent plasmonic chirality is clearly observed in the $k$-space. These findings confirm that the observed phenomenon arises from the collective plasmonic response and its coupling to the handedness of the incident light, rather than from the underlying symmetry.

\section{Conclusion}
In conclusion, we have demonstrated the emergence of non-local chirality from structurally achiral plasmonic hole arrays through collective SPs interactions. Although an individual hole exhibits a helicity-dependent local response via spin–orbit coupling, the periodic repetition of holes enables the plasmonic sources to couple and interfere coherently, producing a collective, Bloch–Floquet-like response. By systematically varying the array size and lattice period, we show that the strength and spatial manifestation of chirality are governed by the combination of the incident beam NA and the structure geometry. Increasing the array size enhances the collective response and progressively reveals the chiral character in both the real and $k$-space. Moreover, the persistence of the chiral response upon changing the lattice symmetry from $C_4$ to $C_6$ demonstrates the robustness of the underlying mechanism. We believe that the role of the Bloch-Floquet effect is crucial for this phenomenon. The investigation of the randomized arrays as an example of uncoupled plasmonic scatterers reveals (in the SI) an absolute suppression of the chirality. We believe that the observed plasmonic chirality has a topological origin however further study is required to address this topic in more detail. In summary, our results establish periodic plasmonic arrays as a platform for converting the hidden chirality of a tightly focused Gaussian beam into a collective non-local chirality of the SP fields. This may provide a general route for generating and controlling chiral plasmonic wavefronts in structurally achiral systems.

\section{DATA AVAILABILITY} 
The data that support the findings of this study are available from the corresponding author upon reasonable request.

\section{Acknowledgment}
Authors acknowledge the Israeli Ministry of Innovation, Science and Technology for funding the research.

\clearpage
\onecolumngrid
\begin{center}
    {\LARGE\bfseries Supplementary Material}\\[2em]
    {\Large\bfseries
    Collective Chiral Response of Achiral Plasmonic Arrays Induced by Truncated Bloch-Floquet Effect
    }
\end{center}
\vspace{2em}
\renewcommand{\thesubsection}{\Alph{subsection}} 
\setcounter{section}{0} 
\renewcommand{\thesection}{S\arabic{section}}
\section{Experimental Setup}
\begin{figure}[ht!]
\centering
\includegraphics[width=1\columnwidth]{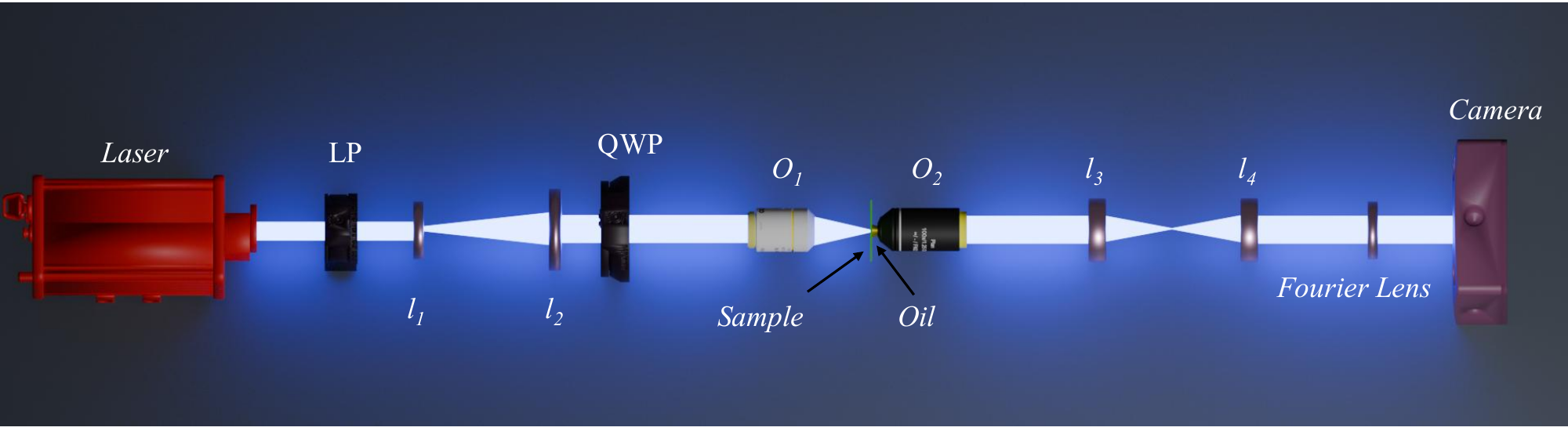}
\caption{\label{Experiment} 
Experimental setup: Scheme of the experimental arrangement showing the illumination path of the laser beam through the Quarter wave plate onto the sample.}
\end{figure}
A linearly polarized Gaussian beam from a diode laser ($\lambda_0=785nm$) was first vertically polarized using a linear polarizer. The beam was subsequently expanded using lenses $l_1$ and $l_2$ to obtain the desired illumination area and then passed through a quarter-wave plate (QWP) to generate right- or left-circularly polarized (RCP/LCP) light. The resulting beam was focused onto the nanostructure using objective $O_1$, with a numerical aperture (NA) of 0.45 and $20\times$ magnification. The surface plasmons (SPs) excited at the sample surface interfere to form the observed plasmonic intensity distribution. The resulting SP field was collected from the glass side using leakage radiation microscopy (LRM) with an oil-immersion objective $O_2$ (NA =1.25, $100\times$ magnification) \cite{drezet2008leakage,hohenau2011surface}. The use of index-matching oil enables the objective to access transverse wavevectors up to its high numerical aperture, including the SP wavevector, $k_{\mathrm{SP}}=k_0\sqrt{\frac{\epsilon_m}{\epsilon_m+\epsilon_d}}$, $k_0=\frac{2\pi}{\lambda_0}$, where \(\epsilon_m\) and \(\epsilon_d\) denote the dielectric permittivities of the gold film and the substrate, respectively. As illustrated in Fig. \ref{Experiment}, lenses \(l_3\) and \(l_4\) form a relay imaging system that projects the collected field onto the camera, enabling direct measurement of the real-space ($r$-space) SP interference pattern. By switching the optical configuration to image the Fourier plane onto the camera, the corresponding momentum-space ($k$-space) distribution was obtained. The plasmonic structure consists of circular holes milled into a $60$ nm-thick gold film deposited on a $160\mu m$-thick glass substrate. The holes of radius 147nm are arranged in a square lattice with different periods as mentioned in the article, these nanostructures were fabricated using focused ion beam milling (Helios 5).

\newpage
\section{Convolution-based modeling of chirality in the $k$-space }

By evaluating the radial wave $C(r) = \frac{e^{ik_{sp}r}}{\sqrt{r}}$ independently, where $r = \sqrt{x^2 +y^2}$, its two-dimensional Fourier transform reveals a highly localized ring in $k$-space. For a purely radially symmetric wave of infinite spatial extent, the dominant spectral weight concentrates precisely at the radial wavevector magnitude $|\mathbf{k}| = k_{sp}$, forming a perfect Dirac delta ring, $\delta\!\left(|\mathbf{k}|-k_{sp}\right)$. However, according to the fundamental scaling properties of the Fourier transform, confining the excitation in $r$-space to a characteristic width $w_0$ mandates a reciprocal spectral broadening in $k$-space. Consequently, the ideal delta ring is convolved with a Gaussian spectral profile, broadening the SP isofrequency contour into an annulus with a finite $k$-space width $\Delta k_r \propto 1/w_0$. Therefore, the fundamental bright annular ring generated by $\tilde{C}$ which is subsequently replicated across the reciprocal lattice by the convolution serves as the direct spectral fingerprint of the SP wavevector, with its observable thickness actively governed by the focal beam waist $w_0$.

\begin{figure}[ht!]
\centering
\includegraphics[width=0.7\columnwidth]{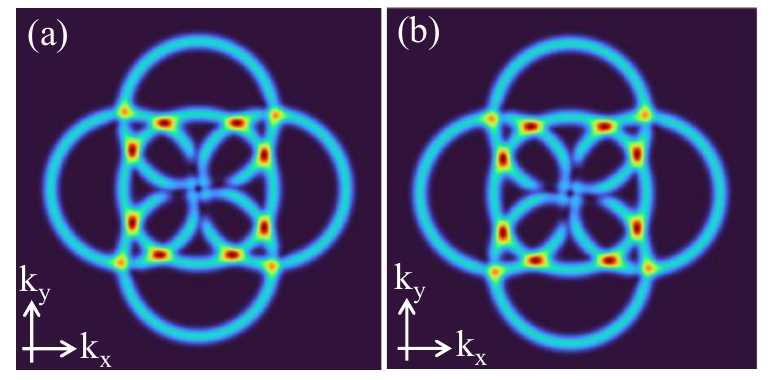}
\caption{\label{Dirac result} (a) and (b) $k$-space representations obtained using the Dirac-delta-ring method for LCP and RCP illumination, respectively.}
\end{figure}

To explicitly visualize how this convolution generates the observed chirality in the $k$-space interference pattern, we can construct a phenomenological model operating entirely within $k$-space. In $k$-space one can model the interference pattern associated with the helical phase of the SP field and convolve it with a cross-shaped lattice of point sources in $k$ space, representing the spatial arrangement of the hole array. The entire formulation lives in $\textbf{k-space}$ $(k_x, k_y)$, where SPs appear as a ring of radius $k_{sp}$. The $k$-space grid is defined as: $k_r = \sqrt{k_x^2 + k_y^2}$, and $\phi = \arctan\!\left(\frac{k_y}{k_x}\right)$. Because SPs exist at a fixed wavevector ($k_{sp}$), their $k$-space signature appears as a Lorentzian-broadened Dirac circle, $\tilde{A}(k_r) = \exp\!\left(-\frac{(k_r - k_{sp})^2}{w_b^2}\right)$, where $w_b$ controls the broadening due to propagation losses and finite source size.  In the lossless limit, $w_b \to 0$ makes $\tilde{A}(k_r) \;\longrightarrow\; \delta(k_r - k_{sp})$ \text{(perfect Dirac circle)}. The phase contributions imprinted on the ring comprise a spiral phase, $\sigma\phi(k_x,k_y)$, and a quadratic phase in $k$-space, which corresponds to a \textit{converging-lens phase} in $r$-space and focuses the SP field toward a focal point. Combining amplitude and phase, the SP ring field in $k$-space is:
\begin{equation}
\begin{aligned}
    \tilde{E}_{\text{ring}}(k_r,\phi) = {} & \exp\!\left(-\frac{(k_r-k_{sp})^2}{w_b^2}\right)  \times \exp\!\Bigl[i\bigl(\sigma\,\phi + \tfrac{k_x^2 + k_y^2}{2k_{sp}} z\bigr)\Bigr]
\end{aligned}
\end{equation}
here we introduce a characteristic length $z$ which scale the conical phase. A cross-shaped lattice of five delta-function point sources is
defined in $k$-space:

\begin{equation}
\begin{aligned}
P(k_x,k_y) ={}&
\delta(k_x,k_y)
+\delta(k_x-k_{sp},k_y)
+\delta(k_x+k_{sp},k_y) \\
&+\delta(k_x,k_y-k_{sp})
+\delta(k_x,k_y+k_{sp}) .
\end{aligned}
\end{equation}

The total $k$-space field is computed as the \textit{convolution} of
the ring with the lattice: $\tilde{E}_{\text{total}}(k_x, k_y) = \tilde{E}_{\text{ring}}(k_x, k_y) * P(k_x, k_y)$. Since convolution with a sum of delta functions replicates and shifts the ring to each lattice point, this expands to:
\begin{equation}
\begin{aligned}
\tilde{E}_{\text{total}} ={}&
\tilde{E}_{\text{ring}}(k_x,k_y)
+\tilde{E}_{\text{ring}}(k_x-k_{sp},k_y) \\
&+\tilde{E}_{\text{ring}}(k_x+k_{sp},k_y)
+\tilde{E}_{\text{ring}}(k_x,k_y-k_{sp}) \\
&+\tilde{E}_{\text{ring}}(k_x,k_y+k_{sp}) .
\end{aligned}
\end{equation}

The observed $k$-space intensity is: $I(k_x, k_y) = \left|\tilde{E}_{\text{total}}(k_x, k_y)\right|^2$, the overlapping rings interfere constructively and destructively, modulated by the phase winding, producing a structured \textbf{$k$-space interference pattern} as shown in Fig. \ref{Dirac result}a,b. This demonstrates that the plasmonic ring carries a winding phase characterized by the sign of $\sigma$. The interference between the rings of the different diffraction orders gives rise to the asymmetric intensity pattern observed around the central diffraction order.

\newpage
\section{Numerical result for hole array with varying size}
\begin{figure}[ht!]
\centering
\includegraphics[width=1\columnwidth]{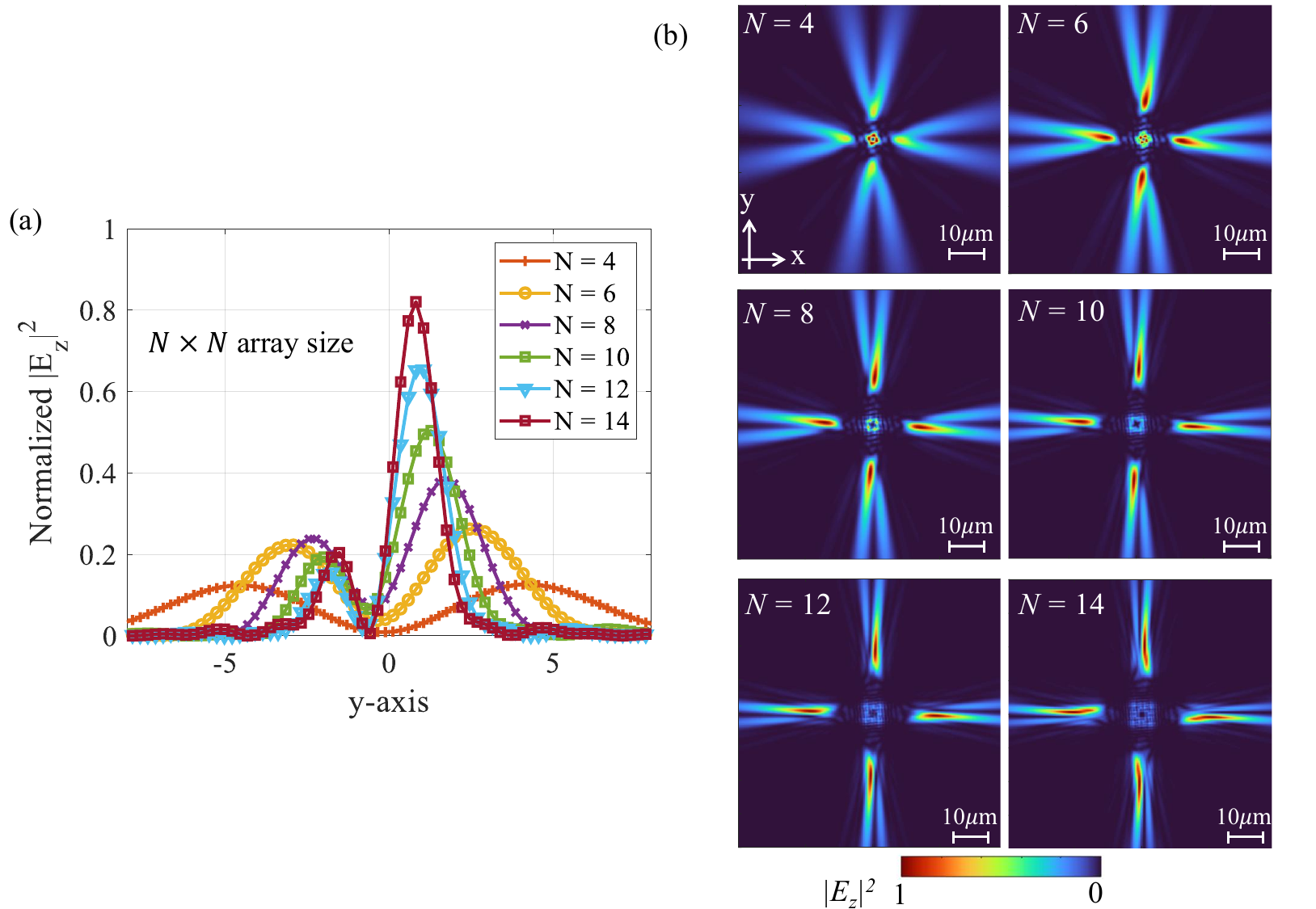}
\caption{\label{numerical result} (a) Intensity distribution along the $y$-axis cross-section. (d) Intensity distribution patterns for different array sizes $N$.}
\end{figure}

The numerical results are presented in Fig. \ref{numerical result}a shows the calculated intensity distribution at a fixed observation plane located at a certain distance from the array. As the array size increases from $N=4$ to $N=14$, the evolution of the plasmonic intensity becomes increasingly pronounced, reflecting the enhanced collective interaction between the holes. The corresponding 2-dimensional intensity distributions for $N=4,6,8,10,12$ and 14 are shown in Fig. \ref{numerical result}b, clearly illustrating the progressive buildup and spatial evolution of the collective plasmonic field with increasing array size.

\begin{figure}[ht!]
\centering
\includegraphics[width=1\columnwidth]{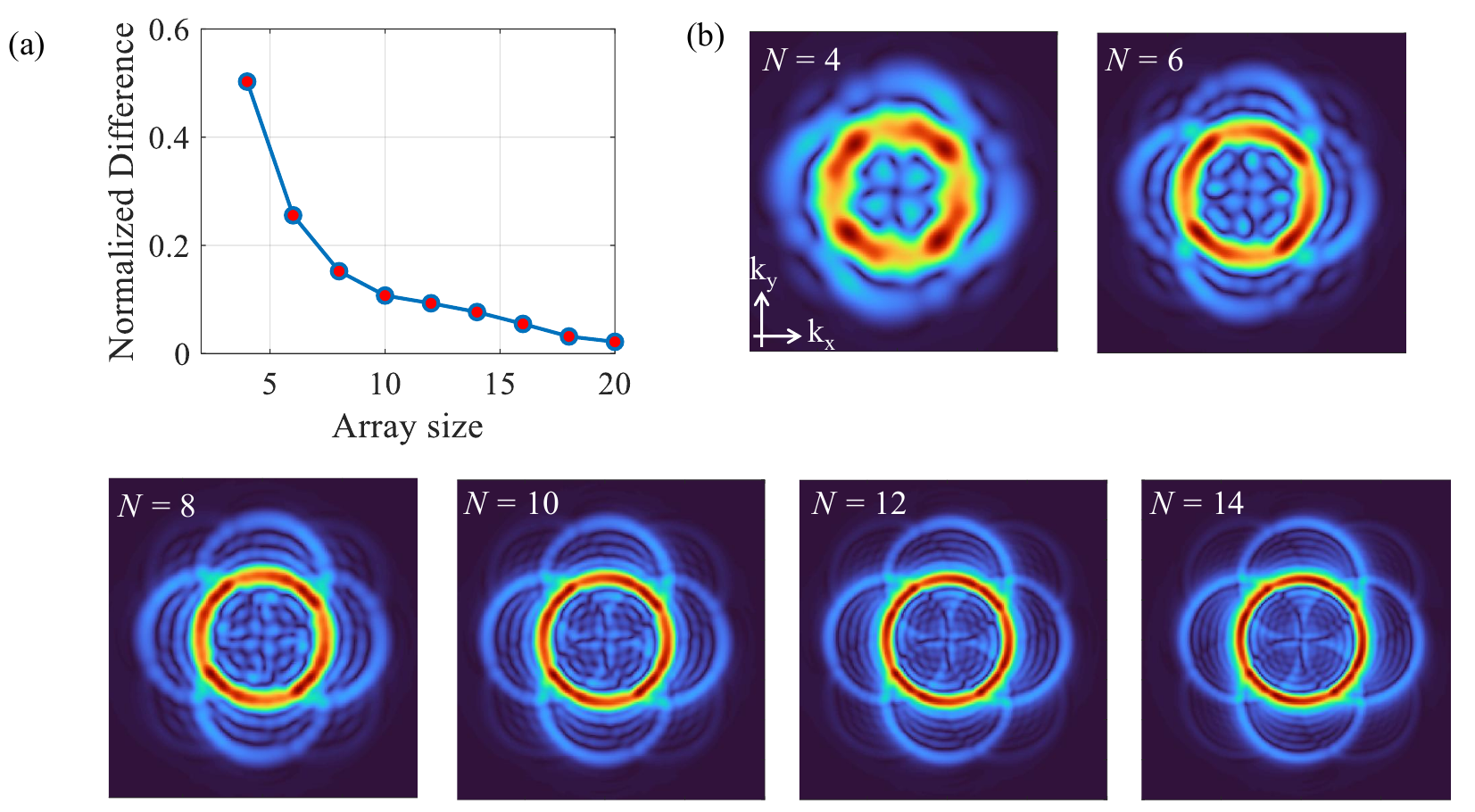}
\caption{\label{Fourier numerical result} (a) Normalized difference as a function of the array size $N$. (b) $k$-space intensity distributions for different array sizes $N$.}
\end{figure}

To quantify the evolution of the $k$-space with increasing $N$ in, we performed a convergence analysis of the calculated $k$-space intensity distributions. For each array size, $k$-space distribution was obtained by applying a two-dimensional Fourier transform to the total field. The convergence was evaluated within a circular region in $k$-space defined by $k_x^2+k_y^2\leq k_{sp}^2,$ where $k_{sp}=2\pi/\lambda_{sp}$ is the free-space wave number. This region provides a common $k$-space domain over which the $k$-space distributions calculated for different array sizes can be compared. For two consecutive array sizes, $N$ and $N-2$, we calculate the normalized difference ($D_N$)

\begin{equation}
\begin{aligned}
D_N=
\frac{
\left\|I_{N}(k_x,k_y)-I_{N-2}(k_x,k_y)\right\|
}{
\left\|I_{N-2}(k_x,k_y)\right\|
},
\end{aligned}
\end{equation}

where $I_N(k_x,k_y)$ denotes the normalized $k$-space intensity distribution of an $N\times N$ array and $|| \cdot ||$ represents the Euclidean norm over the selected circular region. Physically, $D_N$ measures how strongly the $k$-space distribution changes when additional elements are introduced into the array. As the array size increases, an increasing number of elements contribute coherently to the $k$-space, leading to the development of a well-defined collective diffraction structure. Consequently, the addition of further elements produces progressively smaller changes in the $D_N$. When $D_N$ becomes sufficiently small, the $k$-space response can be regarded as effectively converged, meaning that further increasing the array size does not substantially modify the overall $k$-space intensity distribution. This convergence provides evidence that the observed $k$-space structure is a collective property of the periodic array rather than a finite-size effect associated with a particular number of holes. Importantly, the convergence analysis complements the direct visualization of the $k$-space distributions. While the intensity maps show how the collective pattern develops with increasing array size, the normalized difference $D_N$ provides a quantitative measure of when this evolution becomes negligible, as for this reason we have performed till $N = 14$. 

\newpage
\section{Analysis of very large arrays}  
\begin{figure}[ht!]
\centering
\includegraphics[width=1\linewidth]{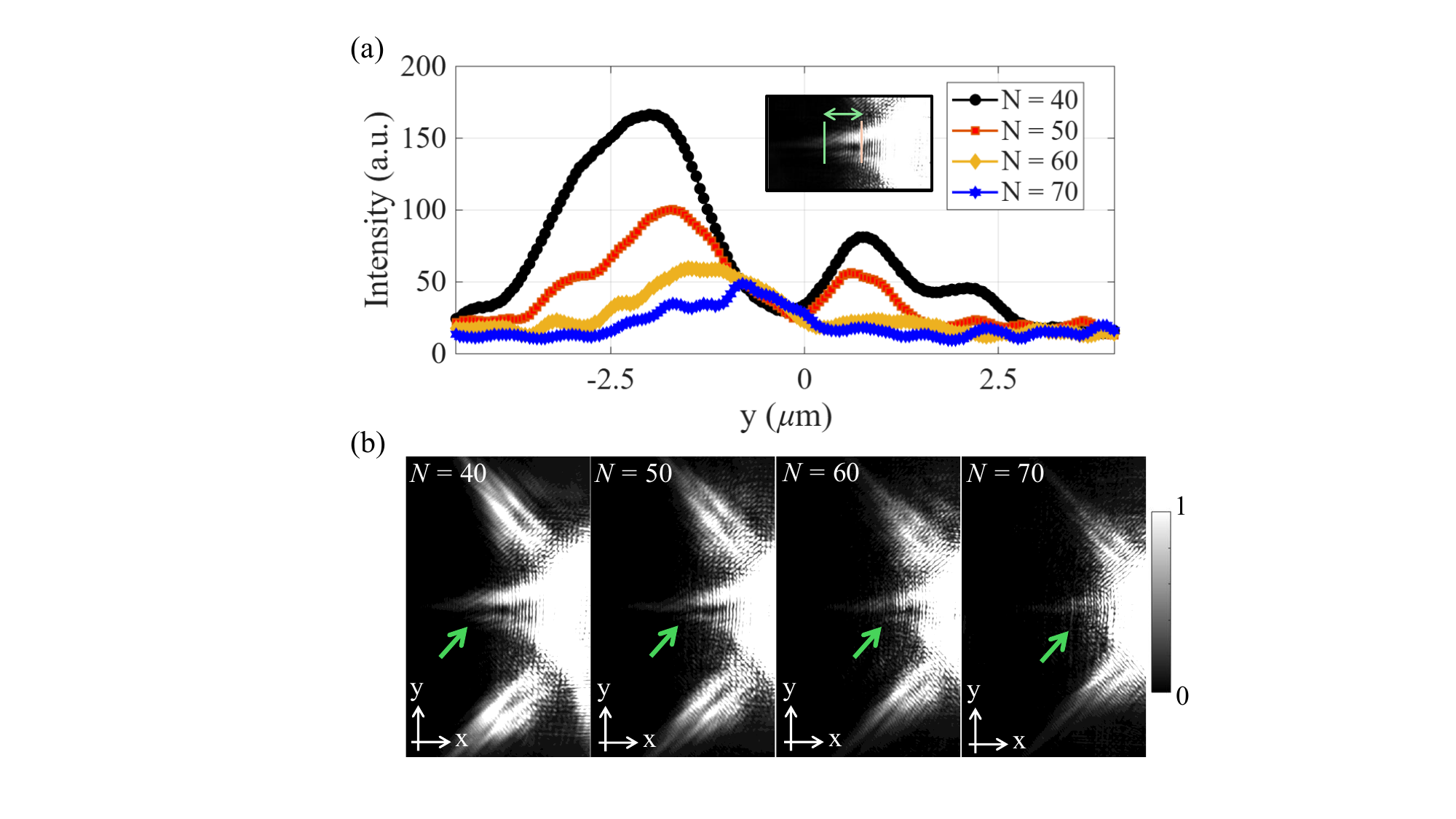}
\caption{\label{BFtheorem} (a) Intensity cross-sections along the $y$-axis for different array sizes $N$. (b) Corresponding intensity distributions from which the cross-sections are extracted.}
\end{figure}

As the lattice size increases, the non-local chiral response is expected to become more pronounced. However, we anticipate that this effect must eventually reach a limit as the SPs propagate beyond the structured array onto the flat Au surface, where no further aperture-induced resonances are available. The resulting field can still interfere with the existing plasmonic waves, producing a helicity-dependent asymmetric intensity distribution. This raises the question of whether the asymmetry persists indefinitely or gradually diminishes as the SPs propagate over the unstructured surface, consistent with the expected constraints of the proposed Bloch-Floquet's theorem. To investigate this, we fabricated hole arrays with different sizes, $N=40,50,60$, and 70. We first fixed the beam size for the $N=40$ array such that the excitation extended to the array edges, allowing the SPs to propagate onto the surrounding flat Au surface. Keeping the beam size unchanged, we then examined the response for progressively larger arrays. As shown in Fig. \ref{BFtheorem}a for RCP illumination, increasing the array size reduces the observed asymmetry, with the intensity distribution becoming increasingly symmetric. The overall intensity also decreases, consistent with the lossy nature of SP propagation. For quantitative comparison, we extracted intensity cross-sections along the $y$-axis and averaged the intensity over multiple cuts along the $x$-axis as shown inset in Fig. \ref{BFtheorem}a. The resulting intensity profiles for different array sizes are presented in Fig. \ref{BFtheorem}b. As the array size increases from $N=40$ to $N=70$, the most distinctive feature is the characteristic dark line associated with the asymmetric intensity distribution gradually disappears, indicating that the asymmetry becomes less pronounced.

\newpage
\section{Analysis of Decoupling Effect in Hole arrays}  
\begin{figure}[ht!]
\centering
\includegraphics[width=1\columnwidth]{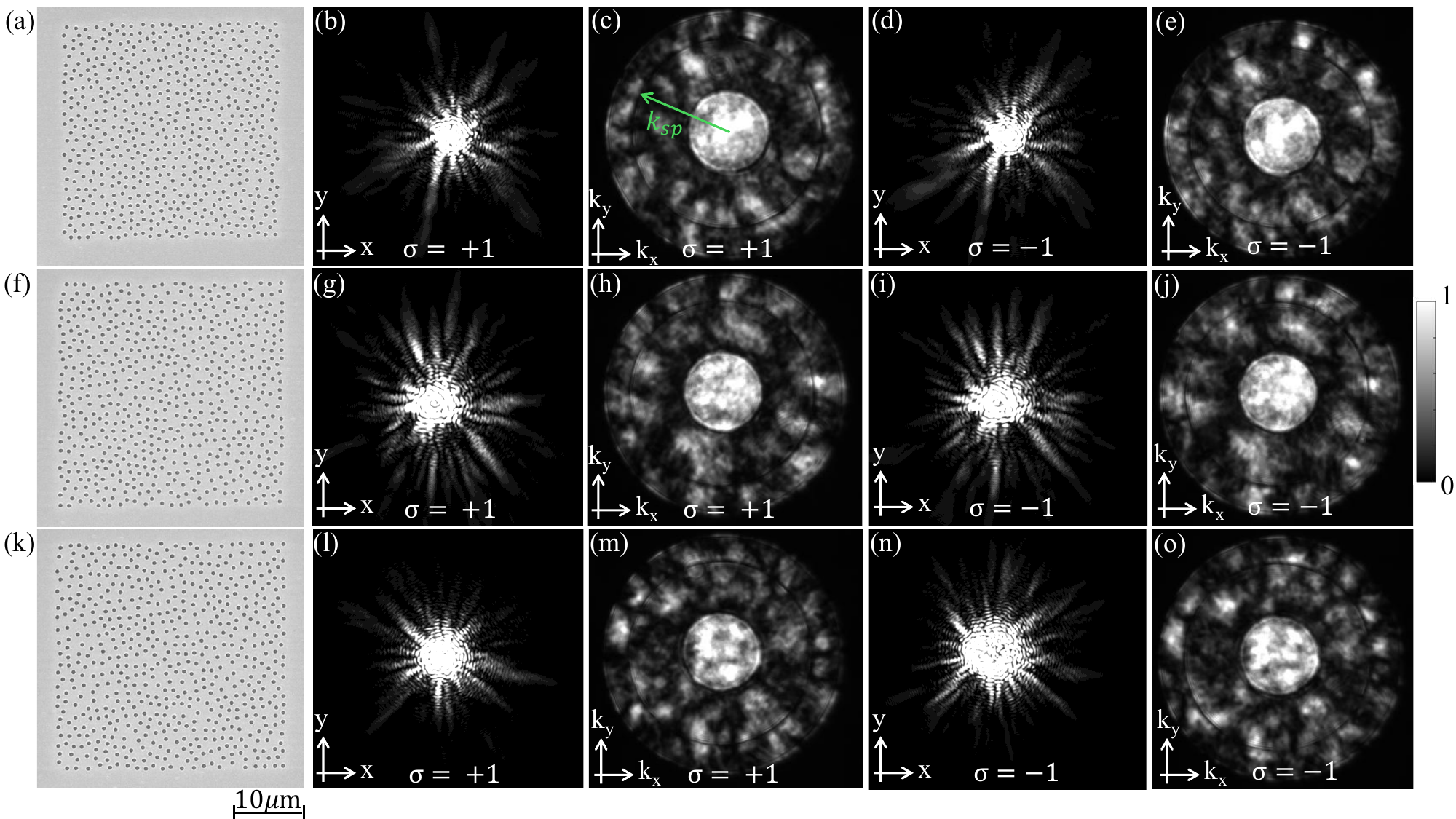}
\caption{\label{decoupling} SEM images of the randomly arranged hole arrays (a,f,k). $r$-space distributions for $\sigma=+1$ (b,g,l) and $\sigma=-1$ (d,i,n). Corresponding $k$-space distributions for $\sigma=+1$ (c,h,m) and $\sigma=-1$ (e,j,o).
}
\end{figure}

To investigate whether the observed $\sigma$-dependent asymmetry originates from the plasmonic coupling between holes or persists independently of their spatial arrangement, we fabricated randomly arranged hole arrays to disrupt the collective coupling of SPs. The arrays were illuminated with CP light of opposite helicities, $\sigma=\pm1$, and their SEM images are shown in Fig. \ref{decoupling}a,f,k. In the $r$-space, both helicities produce scrambled SP distributions, as shown in Fig. \ref{decoupling}b,g,l for $\sigma=+1$ and Fig. \ref{decoupling}d,i,n for $\sigma=-1$. The $k$-space patterns likewise exhibit irregular distributions without distinct diffraction spots. Nevertheless, the plasmonic ring at radius $k_{\mathrm{sp}}$ remains visible, confirming SP excitation on the sample. The corresponding $k$-space patterns are shown in Fig. \ref{decoupling}c,h,m for $\sigma=+1$ and Fig. \ref{decoupling}e,j,o for $\sigma=-1$. This comparison allows us to examine whether the $\sigma$-dependent asymmetry observed in the ordered structures persists when collective plasmonic coupling is disrupted.

\newpage
\section{Measured Results for LCP Illumination of Non-resonant Arrays}
\begin{figure}[ht!]
\centering
\includegraphics[width=0.7\columnwidth]{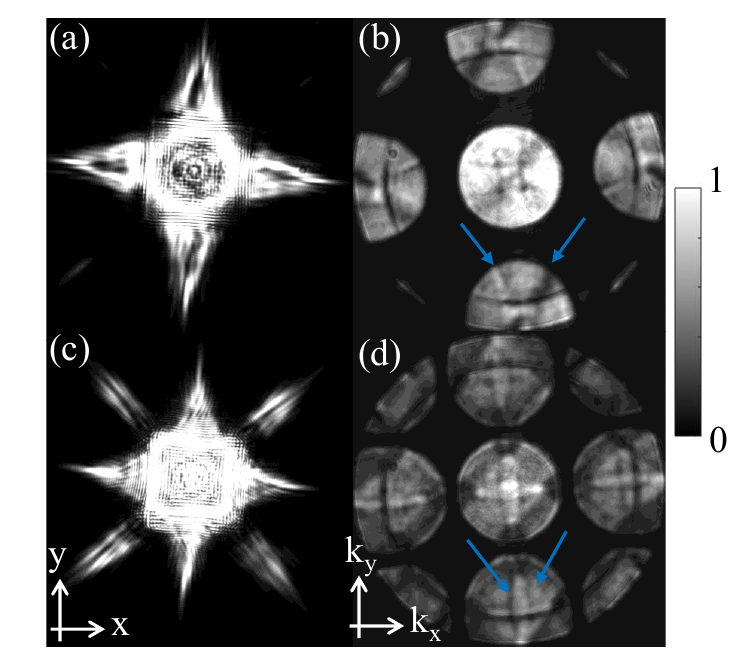}
\caption{\label{Size array}Experimental $r$-space and $k$-space intensity distributions for hole arrays with periods of 705nm a,b and 825nm c,d under RCP illumination.}
\end{figure}

Here, we present the results for LCP illumination of the hole array for two different lattice periods. For a period of $705$ nm, the $r$-space and corresponding $k$-space interference patterns are shown in Fig. \ref{Size array}a and b, respectively. A reduction in the chiral response is observed in both the $r$-space and $k$-space. The same trend is observed when the period is increased to $825$ nm, as shown in Fig. \ref{Size array}c and d, where the $r$-space and $k$-space interference patterns, respectively, exhibit a weaker chiral character. These results further demonstrate the sensitivity of the collective chiral response to the lattice period. 

\newpage
\section{Mesured results for a resonant hexagonal array}
\begin{figure}[ht!]
\centering
\includegraphics[width=0.7\columnwidth]{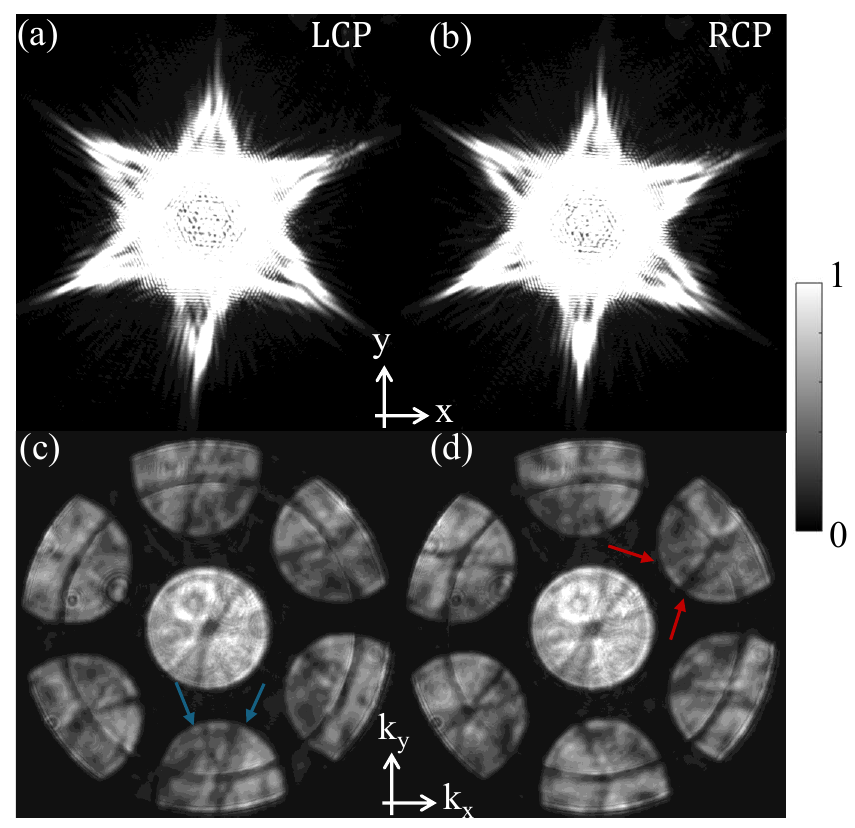}
\caption{\label{Hex Size array} (a,b) $r$-space interference patterns under LCP and RCP illumination, respectively. (c,d) Corresponding $k$-space interference patterns for LCP and RCP illumination.
}
\end{figure}
We have investigated a resonant period of $856$ nm. As shown in Fig. \ref{Hex Size array}c,d, increasing the period leads to a reduction in the chiral response in the $r$-space and produces a characteristic two-line structure in the first-order diffraction spots. The reduced swirling of the plasmonic interference rings further indicates the weakening of the collective chiral response as the inter-aperture distance increases.

\bibliography{apssamp}
\end{document}